\documentclass[a4paper]{article}
\usepackage[a4paper,top=2cm,bottom=2.5cm,left=1.5cm,right=1.5cm,marginparwidth=2cm]{geometry}

\usepackage[english]{babel}
\usepackage{listings}
\usepackage{multirow}

\usepackage{float}

\usepackage{amsmath}
\usepackage{booktabs}
\usepackage[colorlinks=true, allcolors=blue]{hyperref}
\usepackage{listings}
\usepackage{url}
\usepackage{graphicx}
\usepackage{subcaption}

\graphicspath{ {./images/} }

\usepackage{todonotes}
\definecolor{purple1}{rgb}{0.949, 0.941, 0.969}
\definecolor{purple5}{rgb}{0.329, 0.153, 0.561}

\title{An Offer you Cannot Refuse? Trends in the Coerciveness of Amazon Book Recommendations}
\author{Jonathan H. Rystrøm}
\date{June 10th 2023}

\usepackage{apacite}
\usepackage{graphicx}
\usepackage{rotating}
\usepackage{pdflscape}

\begin{document}


\newpage
\maketitle

\begin{abstract}
Recommender systems can be a helpful tool for recommending content but they can also influence users' preferences. One sociological theory for this influence is that companies are incentivised to influence preferences to make users easier to predict and thus more profitable by making it harder to change preferences. This paper seeks to test that theory empirically. We use \textit{Barrier-to-Exit}, a metric for how difficult it is for users to change preferences, to analyse a large dataset of Amazon Book Ratings from 1998 to 2018. We focus the analysis on users who have changed preferences according to Barrier-to-Exit. To assess the growth of Barrier-to-Exit over time, we developed a linear mixed-effects model with crossed random effects for users and categories. Our findings indicate a highly significant growth of Barrier-to-Exit over time, suggesting that it has become more difficult for the analysed subset of users to change their preferences. However, it should be noted that these findings come with several statistical and methodological caveats including sample bias and construct validity issues related to Barrier-to-Exit. We discuss the strengths and limitations of our approach and its implications. Additionally, we highlight the challenges of creating context-sensitive and generalisable measures for complex socio-technical concepts such as "difficulty to change preferences." We conclude with a call for further research: to curb the potential threats of preference manipulation, we need more measures that allow us to compare commercial as well as non-commercial systems. 

\end{abstract}

\vspace{10pt}

\section{Introduction}
What role do recommender systems play in shaping our behaviour? On the one hand, they can seem like a mere convenience: they help us select which music to listen to \shortcite{millecamp_controlling_2018} or which television show to watch \shortcite{bennett_netflix_2007}. When we provide feedback by liking, rating, buying, or interacting with a product, we hope that our actions help the recommender system "learn" our preferences \shortcite{knijnenburg_each_2011}. 

However, what if the recommender systems do not simply \textit{learn} our preferences but \textit{shape} them? By providing recommendations, the recommender systems can influence the products we engage with, which can shape our preferences \shortcite{jiang_degenerate_2019}. This creates a feedback loop that can degenerate into so-called filter bubbles and echo chambers \shortcite{jiang_degenerate_2019}.

The drivers of this could be commercial. The companies behind the recommender systems might have incentives for shaping our behaviour to increase profitability. This is what Zuboff terms the \textit{prediction imperative} in her exposition of \textit{Surveillance Capitalism} \shortcite{zuboff_age_2019}. The prediction imperative states that to secure revenue streams Big Tech companies must become better at predicting the needs of their users. The first step of this is creating better predictive algorithms i.e. going from simple heuristics to sophisticated machine learning \shortcite{raschka_machine_2020}. However, as competition increase, the surest way to \textit{predict} behaviour is to \textit{shape} it \shortcite{zuboff_age_2019}. By shaping behaviour, companies increase predictability at the cost of the users' autonomy \shortcite{varshney_respect_2020}. 

While it may be good business, changing preferences could plausibly count as manipulation. Apart from harming the autonomy of the users \shortcite{varshney_respect_2020}, this could have legal implications under the EU AI Act \shortcite{franklin_missing_2022,kop_eu_2021}.

Take the case of Amazon. In 1998, Amazon introduced item-based \textit{collaborative filtering} \shortcite{linden_amazon_2003} - a simple and scalable recommender model that allows them to recommend similar items. Since then their models have evolved to create more personalised features using machine learning on sophisticated features \shortcite{smith_two_2017}. The effects of this have been more accurate recommendations and - crucially - higher sales \shortcite{wells_amazon_2018}.

The prediction imperative posits that the evolution of Amazon recommender systems should have made it more difficult for users to change preferences to make them more predictable and profitable. They might also steer users towards specific categories that are relatively more profitable \shortcite{zhu_competing_2018}. 

To make such a claim it is essential to have methods for empirically analysing potential manipulation. Fortunately, \shortciteA{rakova_human_2019} provide such a measure: \textit{Barrier-to-Exit}. Barrier-to-Exit provides a measure for how much effort a user must exert to show that they have changed their preferences within a given category. It is built on a theoretical foundation of \shortciteA{selbst_fairness_2019}'s work on fairness traps as well as \textit{systems control theory} as applied to recommender systems \shortcite{jiang_degenerate_2019}. The authors posit that recommender systems with a higher Barrier-to-Exit make it harder to change preferences. 

Methods are ineffective without relevant data to apply them to. This may seem like an insurmountable task: Amazon's recommender system is a complex model that builds on a myriad of advanced features including browsing activity, item features, and buying activity \shortcite{smith_two_2017}. No one but Amazon has access to this data - and they are unlikely to share it \shortcite{burrell_how_2016}. 

We can get around this to some extent by relying on proxies. Specifically, we can use publicly available ratings as a proxy for user input. This has the advantage of being accessible through public datasets \shortcite{ni_justifying_2019}. The disadvantage is that we only have access to a (biased) fraction of the data going into the recommender system.

This paper aims to investigate whether Amazon's recommender system has made it more difficult to change preferences over time. To focus the scope, we will only investigate book recommendations, as books were Amazon's first product \shortcite{smith_two_2017}. This leads us to the following research question:

\begin{itemize}
    \item \textbf{RQ:} Has the Amazon Book Recommender System made it more difficult to change preferences over time?
\end{itemize}

We take several steps to answer this research question. First, we will formalise Barrier-to-Exit in the context of Amazon book recommendations. We will discuss the caveats of the technique and how it relates to preference change. Then we will use a large dataset of Amazon book recommendations \shortcite{ni_justifying_2019} to calculate the Barrier-to-Exit for users who have changed their preferences. We will then analyse the change in Barrier-to-Exit over time using a linear mixed-effects model. Finally, we will discuss the validity and implications of these results.

 This paper has two main contributions to the literature: 1) it provides a novel analysis of trends in preference manipulation in a \textit{commercial} rather than \textit{academic} setting. 2) it assesses the portability of Barrier-to-Exit as a measure for real-world datasets.

\section{Previous Literature}
Several papers have highlighted the need to protect human autonomy and preferences in recommender systems \shortcite{calvo_supporting_2020,varshney_respect_2020,jannach_recommendations_2016}. However, these focus more on the \textit{normative} need to do so, rather than an \textit{empirical} analysis of the phenomenon. With this paper, we aim to fill this gap in the literature.

Most previous empirical analysis of preference manipulation in recommender systems has focused on the MovieLens-dataset \shortcite{harper_movielens_2016}. MovieLens is a movie recommendation platform developed and maintained by the University of Minnesota.\footnote{\href{https://movielens.org}{movielens.org}} The advantages of this dataset are that it is a) rigorously documented as it is maintained by an academic group; b) freely available; and c) well-structured, thus making analysis easier. However, because it is a non-commercial project it is not susceptible to surveillance capitalistic imperatives to the same extent as e.g. Amazon \shortcite{zuboff_age_2019}.

\shortciteA{nguyen_exploring_2014} analysed whether MovieLens users were exposed to less diverse content over time - a type of preference manipulation. While they found a significant (albeit small) decrease, the effect was smaller for users who followed the recommendations than the users who did not. However, as discussed by \shortciteA{rakova_human_2019}, they focus their analysis on highly-active and highly-nonactive users neglecting the middle. Also, content diversity is an important but incomplete measure of preference manipulation. 

\shortciteA{rakova_human_2019} also use the MovieLens dataset to define and showcase Barrier-to-Exit. However, their paper is  more of a proof-of-concept rather than an analysis. The present paper expands on \shortciteA{rakova_human_2019} by applying the metric to a real-world dataset of a commercial recommender system.

\vspace{10pt}
\section{Methods and Data}
\subsection{Defining Barrier-to-Exit} \label{bte}

On a high level, Barrier-to-Exit measures how much effort users must expend to signal that their preferences have changed \shortcite{rakova_human_2019}. It is defined in terms of how quickly users' \textit{revealed preferences} for a specific category change between \textit{interaction thresholds}. In this section, we will motivate the intuition for Barrier-to-Exit as well as formalise the concept within the context of Amazon's recommender system. 

\begin{figure}[H]
    \centering
    \includegraphics[width=0.8\textwidth]{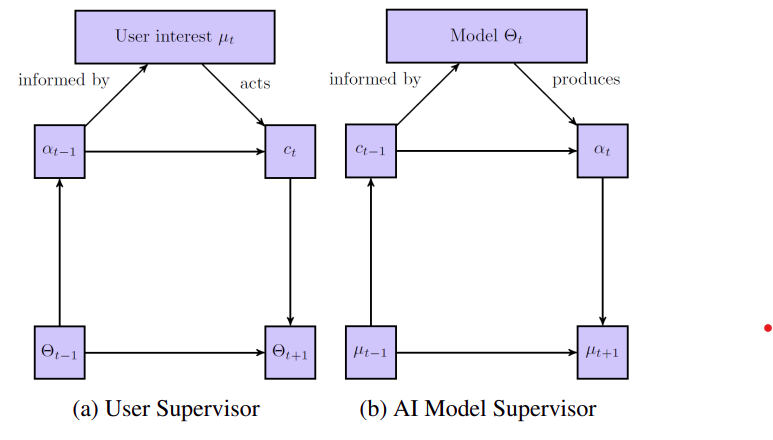}
    \caption{A schematic representation of the control flow in Recommender Systems as seen from the user (a) and AI model (b) perspective. Adapted from \protect\shortciteA{rakova_human_2019}.}
    \label{fig:bte_intuition}
\end{figure}

To understand the role of Barrier-to-Exit and how it can be calculated from ratings, let us consider a diagram of the interaction between the user and recommender system ("AI Model") as seen in Fig. \ref{fig:bte_intuition}. 

Both diagrams (a) and (b) show feedback loops with the user and the model, respectively, as "supervisors". The juxtaposition shows the double-sided interaction as argued in \shortciteA{jiang_degenerate_2019}. The diagram has multiple elements: $\mu$ is user interest, $\Theta$ is the Model, $\alpha$ is the shown recommendations, and $c$ is the revealed preferences (i.e. the signal the model uses to update recommendations). The subscripts denote timesteps going from left to right.

While the diagram acts as a conceptual framework for understanding the interaction, we must consider which parts we can measure and which parts we need to model. \shortciteA{rakova_human_2019} argue that by only analysing how revealed preferences change over time, we can calculate a measure of the effort required to shift preferences; the Barrier-to-Exit. 

Note that while the overall feedback loop concerns the whole model, Barrier-to-Exit is defined per category. Categories can be genres, such as "Thriller" or "Science Fiction", or book types such as "Self-help" or "Cook Book". Each book can have several categories.  

However, we still need to calculate revealed preferences based on ratings. We calculate revealed preference for a given category using ratings and the assigned categories. Following \shortciteA{rakova_human_2019}, let $c^i_t$ be a user's revealed preference for category $i$ at time period $t$. We can calculate this using the following formula:

\begin{equation}
\label{eq:preference}
c^i_t = \sum_{j=1}^{n} m^i_{tj} \cdot r_{tj}
\end{equation}

Here, $n$ is the number of books the user has rated within the time frame; $r_{tj}$ is the rating for book $j$; and $m^i_{tj}$ is the \textit{category-relevance} between the book, $j$, and the category, $i$. Category relevance can be understood as how well a book fits within a given category.

The category relevance is not an automatically available feature of our data (see \ref{section:data}). In contrast, \shortciteA{rakova_human_2019} use the MovieLens dataset \shortcite{harper_movielens_2016}, where category-relevance has been manually annotated for a subset of the data. This makes it possible to use (semi-)supervised learning to annotate the rest of the data (i.e. \shortciteNP{kipf_semi-supervised_2017}).

Unfortunately, the Amazon data has no labels. Instead, we use an unsupervised approach based on category co-occurrence. Books are given a high category relevance for a specific category if they belong to categories that often occur together. For example, a book with the categories "thriller" and "horror" would have a category-relevance score of 1 for "thriller" if it always co-occurs with "horror", but a score of 0 for "gardening" if it never co-occurs with "gardening". We normalize the scores so they range from 0 to 1. See the \href{https://github.com/Rysias/amazon-book-coercion}{GitHub repository} for implementation details.

We now move on to \textit{interaction thresholds} \shortcite{rakova_human_2019}. Conceptually, interaction thresholds are the users' range of preferences within a given category. If, say, a user only ever rates thrillers 4 stars but rates some cookbooks 1 star and others 5 stars, they would have narrow interaction thresholds for thrillers and broader interaction thresholds for cookbooks.

We define the category-specific upper threshold ($X^i_t$) and lower threshold ($Y^i_t$) for a category, $i$, using a temporally rolling window of size $v$ as follows:

\begin{equation} \label{eq:upper_thresh}
    X^i_t = mean(c_{t-v}, \dotsc, c_{t}) + 2*std(c_{t-v}, \dotsc, c_{t}) 
\end{equation}
\begin{equation} \label{eq:lower_thresh}
    Y^i_t = mean(c_{t-v}, \dotsc, c_{t}) - 2*std(c_{t-v}, \dotsc, c_{t}) 
\end{equation}

We denote the average interaction thresholds across all categories as $Y_t$ and $X_t$. Note that time is defined not in discrete steps but in periods. This is because there can be an arbitrary amount of time between two ratings.

Conceptually, the interaction thresholds model the users expected rating behaviour as seen by the model (because of the link going from $c_{t-1}$ to $\Omega_{t}$ in Fig. \ref{fig:bte_intuition}). \shortciteA{rakova_human_2019} posit that the interaction thresholds for non-coercive systems should adapt to make it easier for users to change preferences. 

Finally, we can calculate the Barrier-to-Exit for a given category. This is defined as the sum of preferences that fall between ratings \textit{above} ($t_X$) the thresholds and ratings \textit{below} ($t_Y$) the thresholds. That is: 

\begin{equation}
    BarrierToExit^i_{t_y} = \sum_{\tau \in (t_x, t_y)} c^i_\tau \quad \text{such that} \quad Y_\tau < c^i_\tau < X_\tau
\end{equation}

There are some important things to note about the definition of Barrier-to-Exit. First, there can be multiple values of Barrier-to-Exit per user and category. Every time a user has a preference within a category that goes from above the interaction thresholds to below, a Barrier-to-Exit for that period is defined. 

Second, Barrier-to-Exit defines users who change preferences. Changing preferences are defined as users going from above the interaction thresholds to below the interaction thresholds. 

Third, Barrier-to-Exit cannot be exactly zero. This is because it is only defined when a user has intermediate ratings between the thresholds. If a user has a rating that goes above the interaction thresholds and the next one is below, this would not register in Barrier-to-Exit.

Finally (and crucially), Barrier-to-Exit is only defined for a subset of users. Having a well-defined Barrier-to-Exit for a user requires both a) enough ratings and b) that these ratings change relative to a category. We can thus only draw inferences for this subset of users. We will discuss the implications of this further in the discussion (section \ref{strength-limit}). 

In this section, we have provided a mathematical formulation of Barrier-to-Exit along with important caveats. For the code implementation, please refer to \href{https://github.com/Rysias/amazon-book-coercion}{repository}.

\subsection{Data} \label{section:data}
For this analysis, we use a dataset of Amazon book reviews \shortcite{ni_justifying_2019}. The raw dataset consists of approximately 51 million ratings by ca. 15 million users in the period 1998 to 2018\footnote{For documentation see: https://nijianmo.github.io/amazon/index.html}. All the ratings are on a 1-5 Likert scale. 

The dataset was scraped from the Amazon Web Store building on the methodology of \shortciteA{mcauley_image-based_2015}. Unfortunately, since the dataset lacks a datasheet \shortcite{gebru_datasheets_2021}, it is difficult to figure out whether it has any issues with coverage or bias. It also makes it harder to replicate the data collection from scratch. Other than that, the dataset is easily accessible and well documented.

One coverage-related aspect we need to be aware of is that we are using \textit{ratings} as a proxy for \textit{interactions}. In the dataset, we do not have access to people who bought a product but did not rate it, nor people who neither bought a product nor rated it. This gives us quite an indirect measure of the actual recommendation process - particularly compared to the MovieLens dataset \shortcite{harper_movielens_2016,rakova_human_2019}.

Because of the size of the data, pre-processing becomes non-trivial. An explanation of the necessary steps can be seen in appendix \ref{preprocessing}.

While the original dataset is large, we are only interested in a subset. Specifically, we are interested in users who have changed their preferences. Therefore, we filter to only include users with more than 20 ratings, which follows the conventions in MovieLens \shortcite{harper_movielens_2016} for which Barrier-to-Exit was originally defined \shortcite{rakova_human_2019}. 

\begin{figure}[H]
    \centering \includegraphics[width=0.8\textwidth]{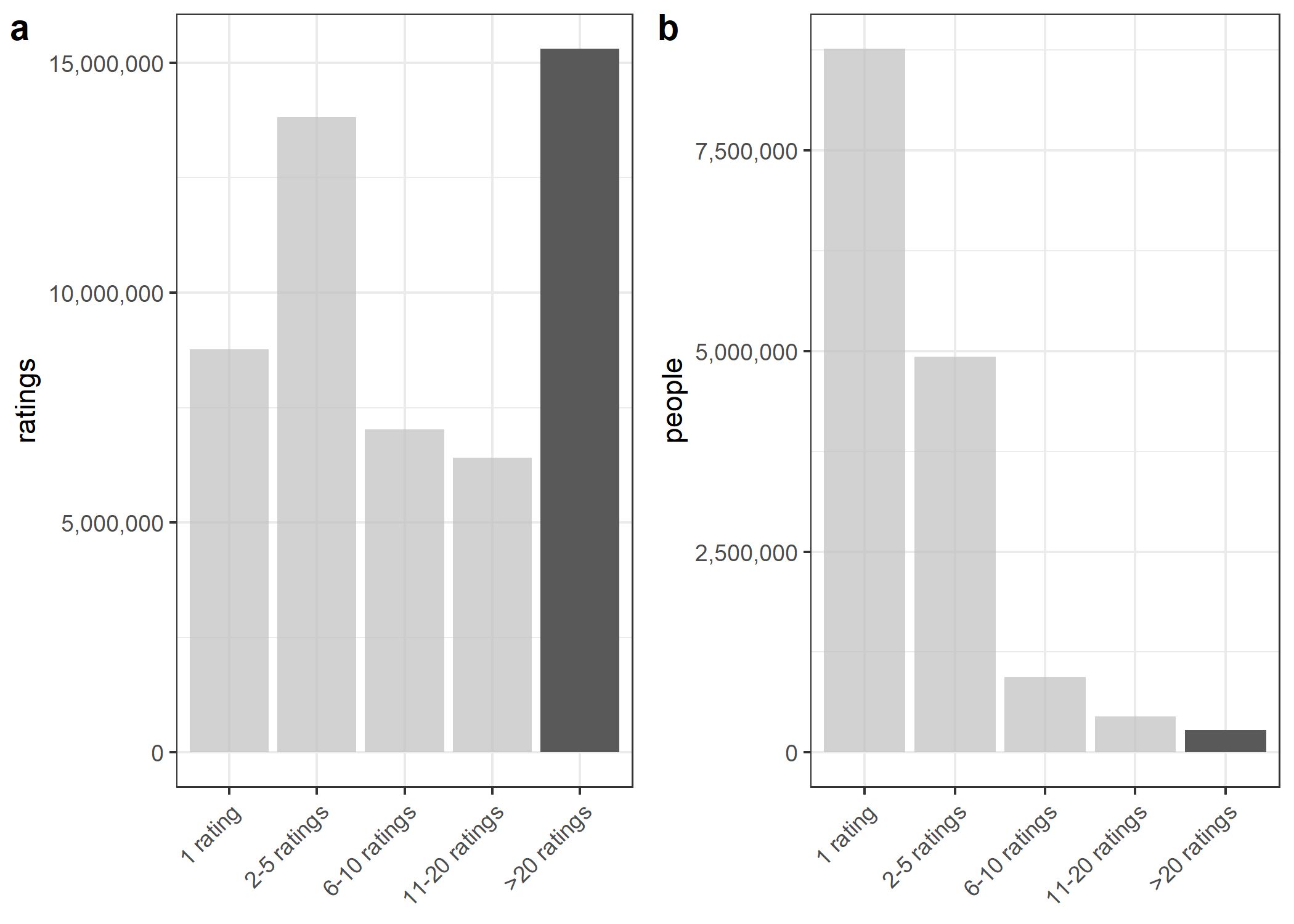}
    \caption{Distribution of total amount of ratings (left) and people (right) across different rating-activity groups. While the group with more than 20 ratings stand for a substantial fraction of the total ratings, they are only a small part of the total users.}
    \label{fig:people}
\end{figure}

Fig. \ref{fig:people} shows the selected subset. It is worth noticing that while our subset retains a substantial fraction of the ratings ($\approx30\%$), we only retain ca. 350,000 users (0.6\%). This is typical for user activity, which tends to be fat-tailed \shortcite{papakyriakopoulos_political_2020}. We will discuss the implications for our interpretation in the discussion (\ref{strength-limit}).

As we will later see, only a fraction of these has changed preferences according to our definition (see section \ref{bte}). For our final analysis, we have 50,626 users which fit our definition ($\approx0.1\%$ of the total).

The rating dataset was merged with a dataset providing categories for each book. The category dataset was from the same source (i.e. \shortciteNP{ni_justifying_2019}). To keep the computations simple for calculating category-similarity (see code on \href{https://github.com/Rysias/amazon-book-coercion}{GitHub}), we only consider categories that have been used on more than 100 books. This approach is valid because the distribution of categories is heavily skewed, meaning that a small number of categories are used on a large number of books. (This is a similar dynamic to user activity; see Fig. \ref{fig:people}). 

\subsection{Model} \label{section:model}
Now that we have operationalised Barrier-to-Exit as a measure of the difficulty to change preferences, let us introduce the statistical model for analysing the trend. 

The first thing to note is that we need a crossed multi-level model \shortcite{baayen_mixed-effects_2008}. Our model should have two levels: user and category. The user level is the most theoretically obvious one. Since each user can have multiple preference changes (with associated Barrier-to-Exit), we should control for their individual differences \shortcite{baayen_mixed-effects_2008}. This is also important as the recommender system will use predictive features that are not accessible in the dataset \shortcite{smith_two_2017}. 

Categories constitute the other level. The role of the category level in our model is to account for item-level features. As explained in the introduction, there are commercial (i.e. companies are following the prediction imperative; \shortcite{zuboff_age_2019}) and algorithmic reasons (i.e. reducing variability could improve on reward objective \shortcite{carroll_estimating_2022}) to believe that different categories will have different Barriers-to-Exit. Categories can therefore act as a proxy for these effects. This crossed design is often used within psychology research \shortcite{baayen_mixed-effects_2008}.

There are two reasons to include categories as random effects and not fixed effects. The first is the number of categories. There are 300+ categories in our dataset. Modelling these as fixed effects would therefore be infeasible. Secondly, since there we use them as a proxy for item-level variance, it is more convenient to only model the random components \shortcite{maddala_use_1971}

This gives us the following model:

\begin{equation} \label{eq:model-equation}
    y_{ij} = \beta_{0} + \beta_{1} X_{1i} + \beta_{2} X_{2i} + e_{i} + e_{j} 
\end{equation}

Here, $y_{ij}$ is the log-transformed Barrier-to-Exit for user $i$ and category $j$; $\beta_0$ is the intercept; $\beta_1$ is the effect of the number of years since the start of the dataset (1998) measured in quarters; $\beta_2$ is the \textit{activity-level}, i.e. the total amount of ratings a user has made in the period of a specific Barrier-to-Exit-measurement; and $e_i$ and $e_j$ are random intercepts for the user and category, respectively. $X_{1i}$ and $X_{2i}$ define the user-level features for time and activity-level, respectively.  

A crucial thing to note is that log-transforming Barrier-to-Exit changes the interpretation of the coefficients. Instead of interpreting them on a \textit{linear} scale, they should be interpreted on a \textit{logarithmic} scale \shortcite{villadsen_statistical_2021}. The most natural way to do this is to exponentiate the effects and interpret it as a percentage change. However, the transformation introduces statistical issues, which we will discuss in section \ref{strength-limit}.

\begin{figure}[H]
\begin{subfigure}{.5\textwidth}
  \centering
  \includegraphics[width=\linewidth]{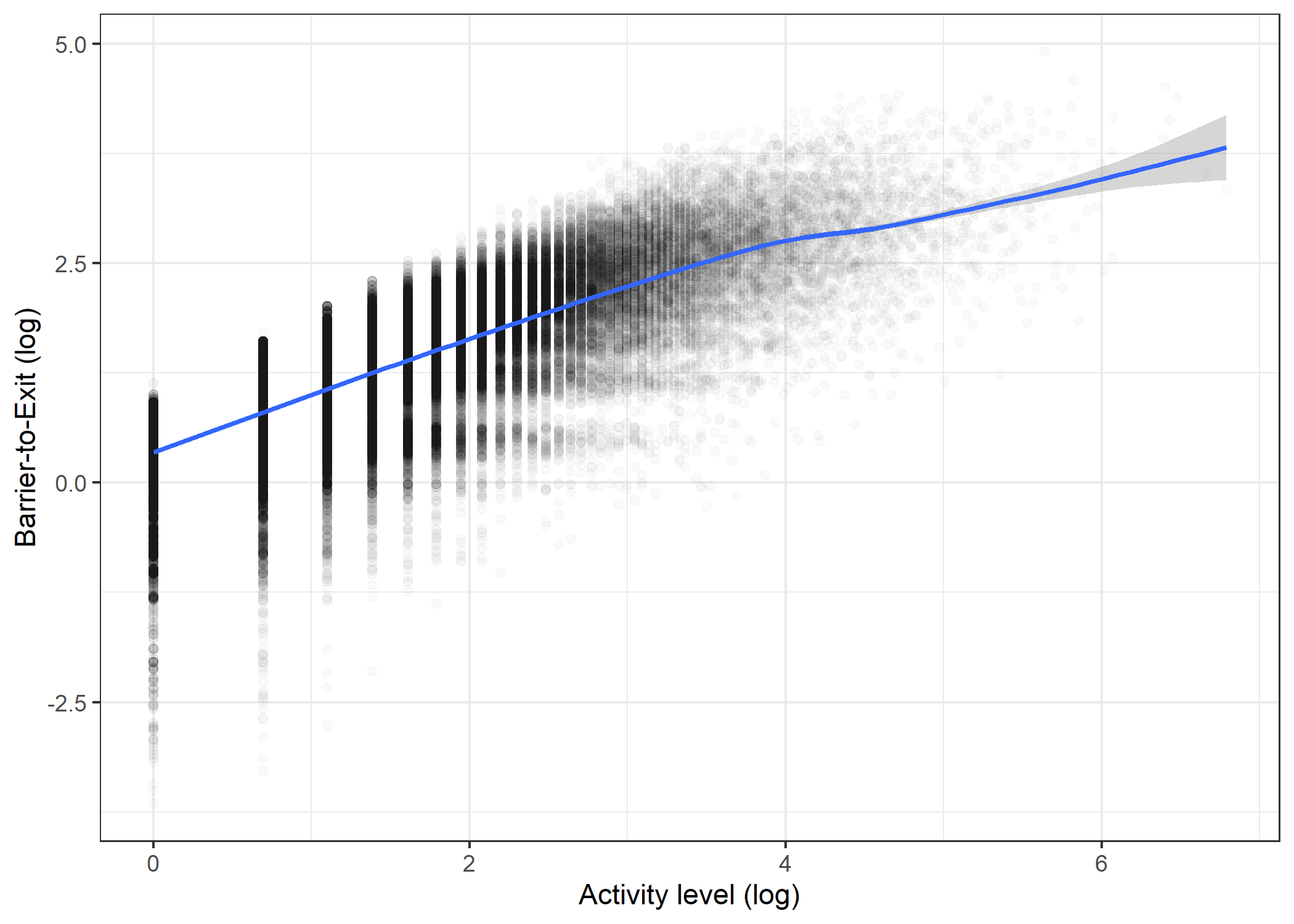}
  \caption{}
  \label{fig:activity-bte}
\end{subfigure}
\begin{subfigure}{.5\textwidth}
  \centering
  \includegraphics[width=\linewidth]{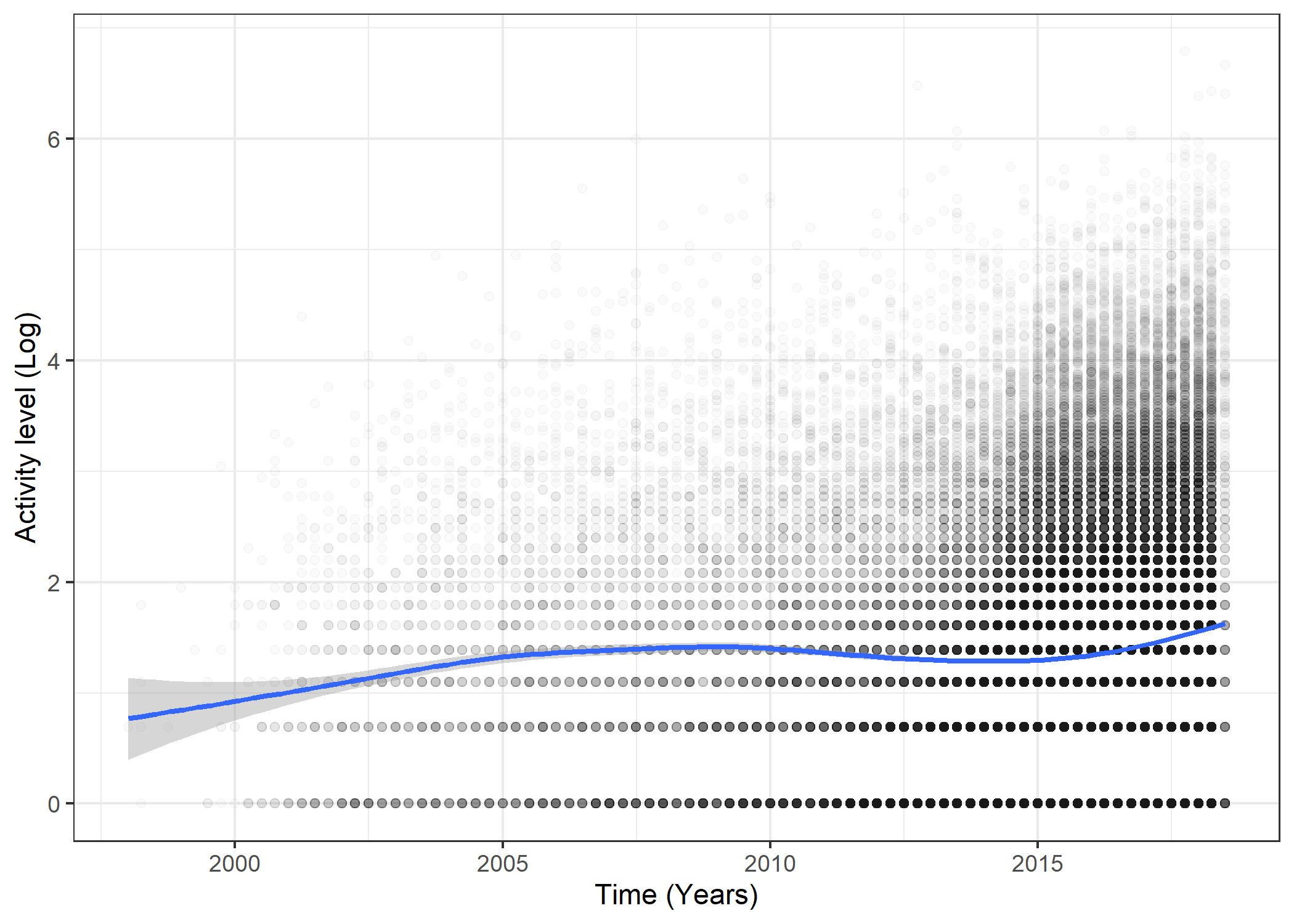}
  \caption{}
  \label{fig:activity-over-time}
\end{subfigure}
\caption{Plots of the activity level, defined as the number of ratings in the period of Barrier-to-Exit. \ref{fig:activity-bte}: The relation between activity-level and Barrier-to-Exit. Notice, the strong linearity. \ref{fig:activity-over-time} Change in activity-level over time. The regression lines are non-parametric GAM regression \protect\shortcite{wood_smoothing_2016}. Note that while there is no strong relation between activity level and time, there has been a substantial increase in the density of points, reflecting Amazon's increased popularity \protect\shortcite{wells_amazon_2018}.}
\label{fig:activity-plots}
\end{figure}

Controlling for user activity level ($\beta_2$) is important. Since Barrier-to-Exit is defined using the \textit{sum} of ratings between the interaction thresholds, users with higher activity levels will tend to have higher Barriers-to-Exit. This can be seen in Fig. \ref{fig:activity-bte}, which shows the relation between activity level and barrier-to-exit, which is increasing. If we do not control for activity level, we introduce trends in the residuals, which invalidates the model (see appendix \ref{no-activity}).

It is also worth noting that activity level is relatively uncorrelated with time (see Fig. \ref{fig:activity-over-time}. This is because activity refers to the activity \textit{within the Barrier-to-Exit} period and not \textit{total activity on Amazon}. The latter has increased substantially as can be seen by the density of the dots in Fig. \ref{fig:activity-over-time}.  

To assess validity, we test the assumptions for the model. For the full check see appendix \ref{assumption}. There are a few violations worth noting: The residuals and random effects deviated from normality - particularly for the category-level random effects. However, this should have little influence on the estimation of the fixed effects \shortcite{schielzeth_robustness_2020}. Nevertheless, we run an additional analysis with the problematic categories removed to assess the robustness of the findings (see \ref{tricky_removed}).

\subsection{Creating and testing hypotheses}
To answer our research question in an inferential framework, we need to transform them into hypotheses with testable implications \shortcite{popper_normal_1970}. We propose the following hypothesis:

\begin{itemize}
    \item \textbf{Hypothesis}: There has been a significant \textit{increase} in Barrier-to-Exit for the Amazon Book Recommender System in the period 1998-2018.
\end{itemize}

To test the hypothesis, we use Satterthwaite’s significance test from the \texttt{lmerTest}-package \shortcite{kuznetsova_lmertest_2017,satterthwaite_approximate_1946} to assess the coefficient for time ($\beta_1$). However, it is important to note that the method of calculating degrees of freedom in mixed effects models \shortcite{satterthwaite_approximate_1946} can inflate Type I errors when the sample size is small \shortcite{baayen_mixed-effects_2008}. In our case, the sample size is large, so this is less of a concern. 

The large sample size also implies p-values close to zero \shortcite{ghasemi_normality_2012} for even small effects. Thus, we are also interested in the \emph{magnitude} of the effect size, rather than just the significance.

Note, that the increase is a \textit{growth-rate} instead of a \textit{linear} increase. This affects how we interpret the magnitude of the effect size.

To assess the model fit we use the marginal and conditional $R^2_{LMM}$ using the \texttt{MuMIn}-package \shortcite{barton_mumin_2022}. This uses an improved method from \shortcite{nakagawa_coefficient_2017}.

\vspace{3ex}
\section{Results}


The results from the model can be seen in table \ref{table:regression_results}. The coefficient for time is 0.018 (SE=0.001). This implies growth in Barrier-to-Exit of 1.8\% per year. This is highly significant (T=29.95, $p \ll 0.0001$). The coefficient for activity level is 0.614 (SE=0.001), which is also highly significant (T=450.11, $p \ll 0.0001$). 

\begin{table}[!htbp] 
  \centering 
  \caption{Main Results} 
  \label{table:regression_results} 
  \begin{tabular}{lc} 
    \toprule
    & \textit{Dependent variable:} \\
    \cmidrule(lr){2-2}
    & Barrier-to-Exit (log) \\ 
    \midrule
    Time (years) & 0.018$^{***}$ \\ 
    & (0.001) \\ 
    \addlinespace
    Activity-level (Log) & 0.613$^{***}$ \\ 
    & (0.001) \\ 
    \addlinespace
    Intercept & 0.300$^{***}$ \\ 
    & (0.022) \\ 
    \midrule
    Unique Users (VPC) & 51,208 (0.24) \\
    Unique categories (VPC) & 330 (0.34) \\
    Observations & 84,806 \\
    \midrule
    Marginal $R^2$ & 0.64 \\
    Conditional $R^2$ & 0.81 \\
    \bottomrule
    \multicolumn{2}{r}{\textit{Note:} $^{*}$p$<$0.1; $^{**}$p$<$0.05; $^{***}$p$<$0.01} \\
  \end{tabular} 
\end{table}

The marginal $R^2$ is 0.64, i.e. 64\% of the variance is explained by the fixed effects. The conditional $R^2$ is 0.81, i.e. 81\% of the variance is explained by the random and fixed effects combined. For the random effects, the user-level VPC is 0.24 and the category-level VPC is 0.34. These should be interpreted cautiously as discussed in Section \ref{strength-limit}.

\begin{figure}[H]
\begin{subfigure}{.5\textwidth}
  \centering
  \includegraphics[width=\linewidth]{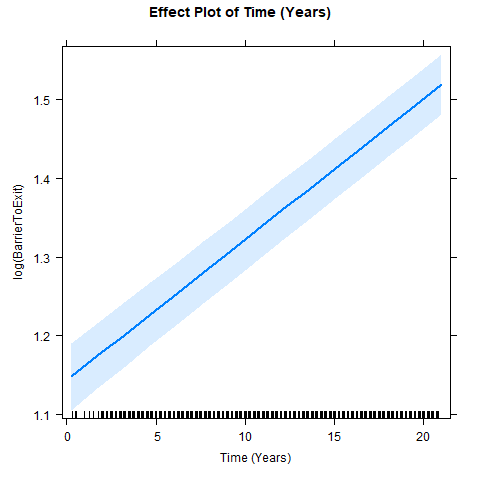}
  \caption{}
  \label{fig:effect_plot_years}
\end{subfigure}
\begin{subfigure}{.5\textwidth}
  \centering
  \includegraphics[width=\linewidth]{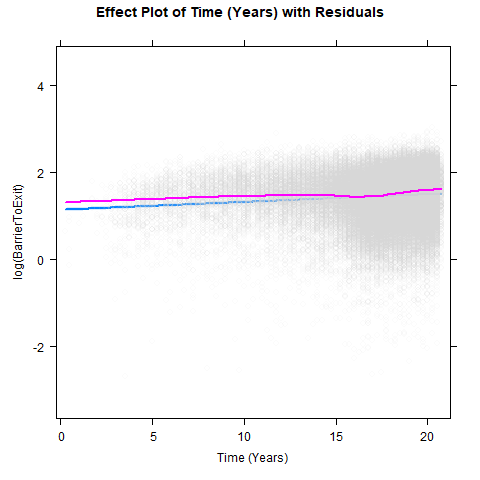}
  \caption{}
  \label{fig:effect_plot_years_resids}
\end{subfigure}
\caption{Effect plots for Year. \ref{fig:effect_plot_years} shows the partial effect plot. \ref{fig:effect_plot_years_resids} shows the same but with residuals added. The X-axis show years after the beginning of the dataset (1998 = 0). The pink line in Fig. \ref{fig:effect_plot_years_resids} is a non-parametric line of best fit.}
\label{fig:year_effects}
\end{figure}

A visual representation of these models can be seen in figure \ref{fig:year_effects}. The partial effects plot \shortcite{fox_effect_2003} in fig \ref{fig:effect_plot_years} shows an increase in Barrier-to-Exit from approximately 1.15 to 1.5. This translates into a growth of approximately 43\% over the duration of the study. Fig. \ref{fig:effect_plot_years_resids} shows the effect plot with residuals. 

\vspace{3ex}
\section{Discussion}
\subsection{Key Findings}
Recall our research question:
\begin{itemize}
    \item \textbf{RQ:} Has the Amazon Book Recommender System made it more difficult for users to change preferences over time?
\end{itemize}

Our analysis finds highly significant growth in Barrier-to-Exit over time in the period spanning from 1998 to 2018. We can therefore reject the null hypothesis of no change, which provides evidence that the Amazon Book Recommender system has indeed made it more difficult to change preferences. 

The growth rate of Barrier-to-Exit is approximately 1.8\%. This implies that over the 20-year period of the dataset, the Barrier-to-Exit has increased by approximately 43\%.\footnote{$1.018^{20} \approx 1.43$} While extrapolating is a tricky matter - particularly, for non-linear parameters \shortcite{timmers_inverse_1977}. However, we must moderate these findings in light of the limitations of the study, which discuss in the next section.

\subsection{Strengths and Limitations} \label{strength-limit}
There are both strengths and weaknesses in our analysis. The main strength is the scale of the analysis. This scale allowed us to test for the relatively small effect sizes - even while relying on noise-introducing approximations like defining category-relevance using co-occurrence (see \ref{bte}). The scale also allowed us to test the feasibility of using Barrier-to-Exit in practice, which we will return to later. 

However, the design also has some significant limitations, particularly with a) the validity of the proxy (i.e. the \textit{construct validity}) and b) the potential sample bias. 

The primary issue with the validity of the proxy is that we have an \textit{indirect} and \textit{incomplete} view of the recommendation process. The ratings only constitute a small part of the interaction with the recommender system. The primary feedback loop is plausibly in the purchasing and browsing behaviour \shortcite{smith_two_2017}. This breaks with the original framing from \shortcite{rakova_human_2019}, which assumes a more direct interaction between ratings and recommender systems. While this assumption holds for MovieLens \shortcite{harper_movielens_2016}, it is more problematic for Amazon. The MovieLens recommender system is made around ratings; the "contract" between the user and MovieLens is that the user provides ratings and MovieLens gives personalised recommendations based on those ratings. 

The ratings on Amazon, however, have a more public purpose: they help other consumers choose products \shortcite{leino_case_2007}. In some sense, this makes ratings a strong signal for preferences \shortcite{leino_case_2007}. 

The main implication of the validity is thus one of coverage. The Amazon rating process thus provides a cost both in terms of time (you have to publicly create a review) and money (most users probably rate products they bought; \shortciteNP{leino_case_2007}). Thus, the average Amazon user in the filtered dataset made 43 ratings while the average MovieLens user has made ca. 740 ratings \shortcite{harper_movielens_2016}.\footnote{Bear in mind the long-tailed nature of both distributions.}

This leads us to potential sample bias. Because Barrier-to-Exit requires relatively many reviews to have a well-defined value, our analysis has predominantly very active users. This introduces a bias: we can only draw inferences for this particular subset of users and not the general population of Amazon customers. The sample bias shows a problematic aspect of Barrier-to-Exit as a model for preference change, which we will discuss further in section \ref{section:implications}. 

These active users plausibly represent a substantial fraction of Amazon's revenue. However, though there is plausibly a correlation between the number of ratings a user has made and commercial interest for Amazon, this need not be the case. 

Additionally, there are important limitations with the statistical analysis that must be addressed. 

First, some assumptions were violated (see appendix \ref{assumption}). Specifically, this has to do with the normality of the residuals and random effects. These violations arise from ill-behaving categories (see \ref{fig:qq_residuals_main} in appendix \ref{assumption}). Theoretically, this should not affect the fixed effects estimates \shortcite{schielzeth_robustness_2020}. Accordingly, re-fitting the model with the problematic categories removed showed similar results (see appendix \ref{tricky_removed}). However, it does affect the interpretation of the VPCs of the two categories. Since these are mainly used to control for variability in the levels \shortcite{baayen_mixed-effects_2008} and not for testing hypotheses (like described by \shortciteNP{maddala_use_1971}), this is a minor problem. 

There are two potential statistical causes of these issues. The first is a lack of data from the early years of Amazon. As fig. \ref{fig:effect_plot_years_resids} shows, there are very few observations of Barrier-to-Exit in the early years compared to later. Amazon has grown dramatically in the past 25 years \shortcite{wells_amazon_2018}, which has been fuelled by many more people having access to the internet \shortcite{pandita_internet_2017}. Statistically, this makes it more difficult to assess the long-term increase as early observations will tend to have high leverage \shortcite{fox_applied_2015}. 

This leads us to the second issue: problems with transformations. Transforming the data introduces problems for the validity and interpretability of the results \shortcite{feng_log-transformation_2014}. This makes some researchers argue that it is better to refrain - even when the assumptions are violated \shortcite{schielzeth_robustness_2020}. 

In our case, fitting the models without transforming the Barrier-to-Exit made it impossible to fit the models without singular fits \shortcite{fox_applied_2015}. Since our data had no zero-values and had a relatively high mean given the skew, we could log the data while avoiding the most serious problems from the transformation \shortcite{ohara_not_2010,ghasemi_normality_2012}. 

However, there are alternatives. We could have utilised \textit{generalised linear mixed models} (GLMM; \shortciteNP{fox_effect_2003}). GLMMs allow us to specify a link function which makes it possible to account for the non-normality in a more elegant way \shortcite{fox_applied_2015}. As Barrier-to-Exit is continuous and right-skewed it might be well-fitted by a gamma-distribution \shortcite{nakagawa_coefficient_2017}. As a \textit{post-hoc} test, we fit a gamma GLMM to the data (see \ref{gamma model}). The findings are similar (positive growth in Barrier-to-Exit), however, problems with the fit prohibit strong conclusions. This highlights the added complexity of modelling with GLMMs \shortcite{fox_effect_2003}.

\subsection{Implications and Further Research} \label{section:implications}
Because of the issues discussed above, we fail to provide a "smoking gun" for preference manipulation in Amazon's book recommender system. Nevertheless, in attempting to model the evolution of Barrier-to-Exit for Amazon, we have uncovered some important perspectives that can inform further work in investigating preference manipulation. 

First, it is essential to have metrics that are defined for the entire population. Metrics can exclude certain people either in their definition or execution. As previously discussed, Barrier-to-Exit is only defined for people who have made several ratings about the same category within the specified time window. This excludes both customers who choose not to (publicly) rate products and customers who only use Amazon sporadically. Previous research suggests that "lurkers" (people who do not actively post) make up a substantial part of the internet population \shortcite{nonnecke_lurker_2000}. Investigating preference manipulation for these users is important - both ethically \shortcite{jannach_recommendations_2016} and legally \shortcite{franklin_missing_2022}. However, with Barrier-to-Exit we lack the data to accomplish this.

One way to get broader coverage is to shift the metrics from the \textit{user-level} to the \textit{system-level} - i.e. whether the recommender system manipulates preferences in general. In this paper, we use the aggregate Barrier-to-Exit of many users to investigate the trend over time. By instead focusing on the system itself we could expand our toolbox. This includes \textit{"sock puppet"-auditing} \shortcite{sandvig_auditing_2014}: creating fake profiles to interact with recommender systems in controlled ways. "Sock puppet"-audits have been used to investigate whether different recommender systems facilitate radicalization \shortcite{ledwich_algorithmic_2019}. However, the methodology comes with its own set of practical and ethical limitations (see \shortciteNP{sandvig_auditing_2014}). 

Second, there is a dilemma of \textit{portability} (i.e. how well the metric can be used across contexts; see \shortciteNP{selbst_fairness_2019}). On the one hand, socio-technical metrics (like Barrier-to-Exit) need to be tailored to their context. Blindly "porting" metrics from one domain to another can obscure the original purpose. On the other hand, portability between different systems is necessary for comparison. 

Barrier-to-Exit was designed for a content recommendation system based on user ratings. Intuitively, that should make it well-suited for use on Amazon book recommender; The setup of the data is similar \shortcite{harper_movielens_2016,ni_justifying_2019}. Nevertheless, the difference in context makes it difficult to port Barrier-to-Exit to Amazon: this introduces the issues with sampling bias, which we discussed earlier.

One solution is to rely on audits conducted by the companies themselves. These audits would provide a more accurate estimate of user preferences than external ratings as they have access to proprietary data. However, this raises ethical concerns as the companies may not accurately report any negative findings about their systems. Some researchers support this type of self-governance \shortcite{roski_enhancing_2021}, while others are sceptical \shortcite{zuboff_age_2019}. In any case, there must be mechanisms in place to verify these audits in order to establish trust and comply with regulations \shortcite{floridi_capai_2022}.

Further work should focus on creating measures of preference manipulation in content-based recommender systems. These should focus on having a high-construct validity and a high coverage (i.e. measure "actual" effort of preference change for close to all users).

\vspace{3ex}
\section{Conclusion}
Understanding how recommender systems shape our behaviour is essential to avoid manipulation. In this paper, we investigated the Amazon recommender system concerning whether it has made it harder to change preferences. By analysing the Barrier-to-Exit \shortcite{rakova_human_2019} of more than 50,000 users, we found a highly significant growth in Barrier-to-Exit over time, which indicates that it has indeed become harder to change preferences for the analysed users.

However, sampling bias induced by the calculation of Barrier-to-Exit makes it difficult to draw conclusions about the general population of Amazon customers. This highlights the dilemma of portability in measuring socio-technical systems \shortcite{selbst_fairness_2019}: accurately evaluating a concept like "changing preferences" requires adapting to the context of the system, which makes it more difficult to generalise (and compare) to other systems. 

Comparing recommender systems is necessary for ensuring that these respect human autonomy \shortcite{varshney_respect_2020} and live up to new regulations such as the EU AI Act \shortcite{kop_eu_2021}. Further work, should aim to create auditing procedures and metrics that allow third parties to measure potential preference manipulation in a way that fits within the context of the industry \emph{and} allows for comparisons between different systems. This will help assess the pressures of \emph{Surveillance Capitalism} \shortcite{zuboff_age_2019} on human autonomy. 


\bibliographystyle{apacite}
\bibliography{new_references}


\appendix

\section {Validation of Assumptions} \label{assumption}
There are several assumptions to validate for crossed-linear mixed effects models \shortcite{baayen_mixed-effects_2008}. Specifically, we need to assess:
\begin{itemize}
    \item \textbf{Linearity}: Is the phenomenon the model captures actually linear? This can be tested by looking for trends in the residuals. If no trends are found, linearity holds \shortcite{poole_assumptions_1971}.
    \item \textbf{Homogeneity of Variance}: Is the variance across residuals equal? This can be tested by plotting residuals against response and looking for cone-like shapes \shortcite{fox_applied_2015}. 
    \item \textbf{Normality of residuals}: Are the residuals normally distributed? We can assess this by a QQ-plot over the residuals. 
    \item \textbf{Normality of random effects}: Per the definition of crossed-effects mixed effects models \shortcite{baayen_mixed-effects_2008}, we expect the residuals to be normal. This can be assessed similarly to the random effects.
\end{itemize}

Note that testing the assumptions relies on \textit{visual} tests rather than \textit{statistical} tests. This is not for a lack of statistical tests (see e.g. \shortciteNP{fox_applied_2015}). Rather it is because statistical tests tend to become oversensitive for large datasets \shortcite{ghasemi_normality_2012}. Given our dataset has more than 50,000 observations, it is safer to use visual tests. 

First, let us test for linearity and heteroscedasticity. The residual plot can be seen in fig. \ref{fig:residuals_main}:

\begin{figure}[H]
    \centering
    \includegraphics[width=0.5\textwidth]{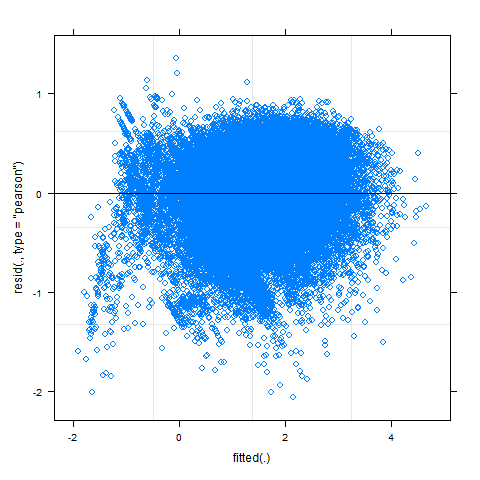}
    \caption{Residuals for the main model}
    \label{fig:residuals_main}
\end{figure}

At first glance, there seems to be no strong heteroscedasticity or apparent linearity. However, at a closer look, there are some weird outliers in the bottom left corner. Furthermore, there is a weird kind of straight line going from the top left corner and down towards the middle. This indicates that there may be something wrong with the model.

Let us now assess the normality of the residuals:

\begin{figure}[H]
    \centering
    \includegraphics[width=0.5\textwidth]{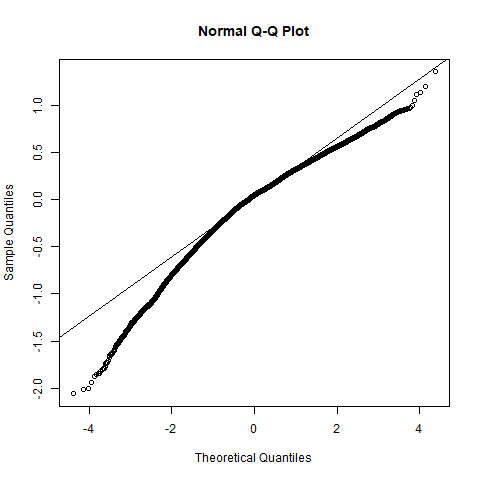}
    \caption{Caption}
    \label{fig:qq_residuals_main}
\end{figure}

We notice that the residuals are generally lower than the normality line. This indicates a left skew in the residuals, i.e. there are smaller outliers. While the fixed effects are robust against this type of skew, the estimates for group-level variances are more affected \shortcite{schielzeth_robustness_2020}.

Finally, let us consider the normality of the random effects. A qq-plot of the random effects for the two levels can be seen below:

\begin{figure}[H]
    \centering
    \includegraphics[width=0.5\textwidth]{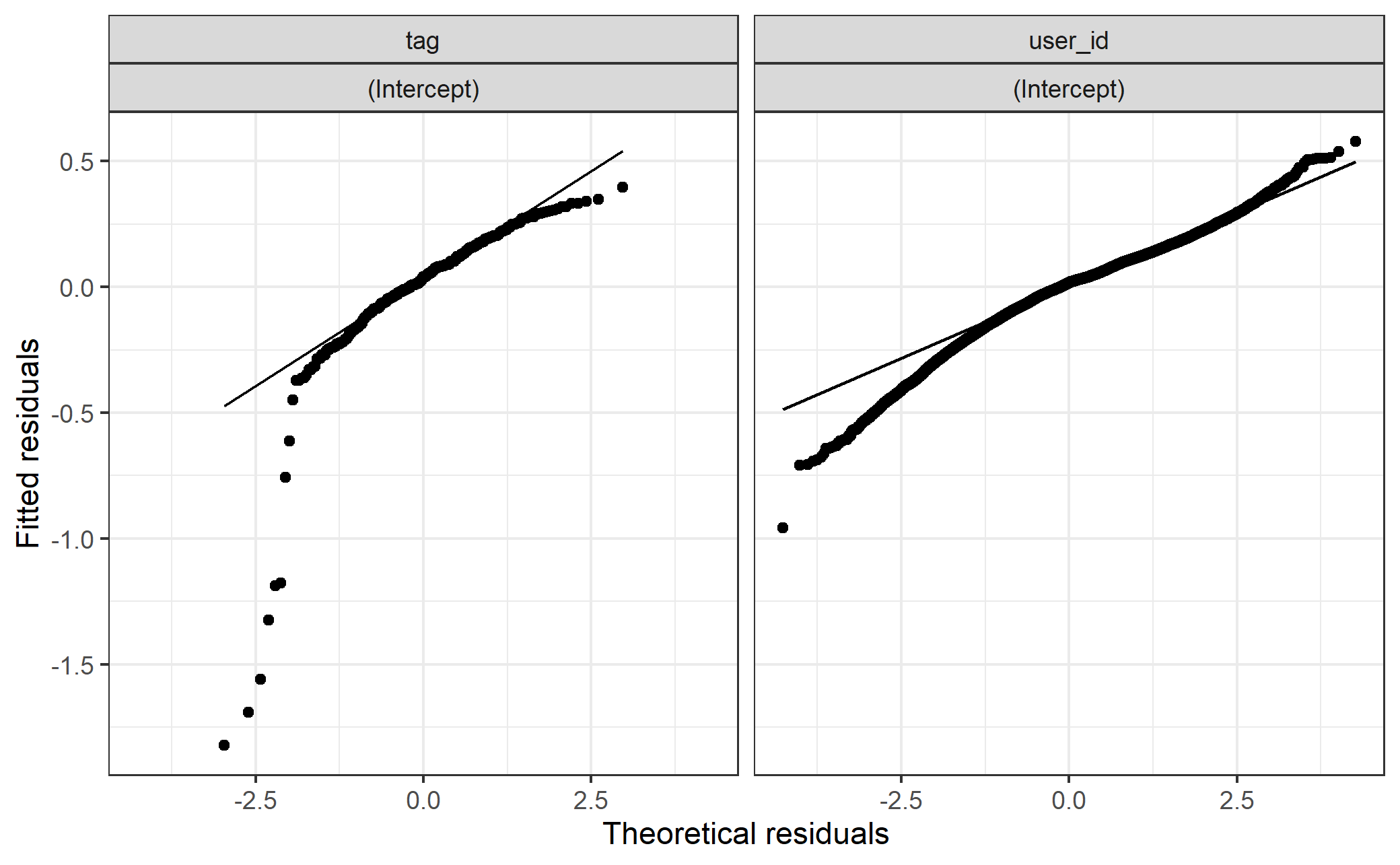}
    \caption{QQ-plot of the random effects per group. "tag" represents categories and "$user\_id$" represents users}
    \label{fig:qq_main_random}
\end{figure}

Here we see a dramatic skew toward the lower level of the fitted residuals. This implies that there are weird dynamics in the lower values of Barrier-to-Exit.

\begin{figure}
    \centering
    \includegraphics{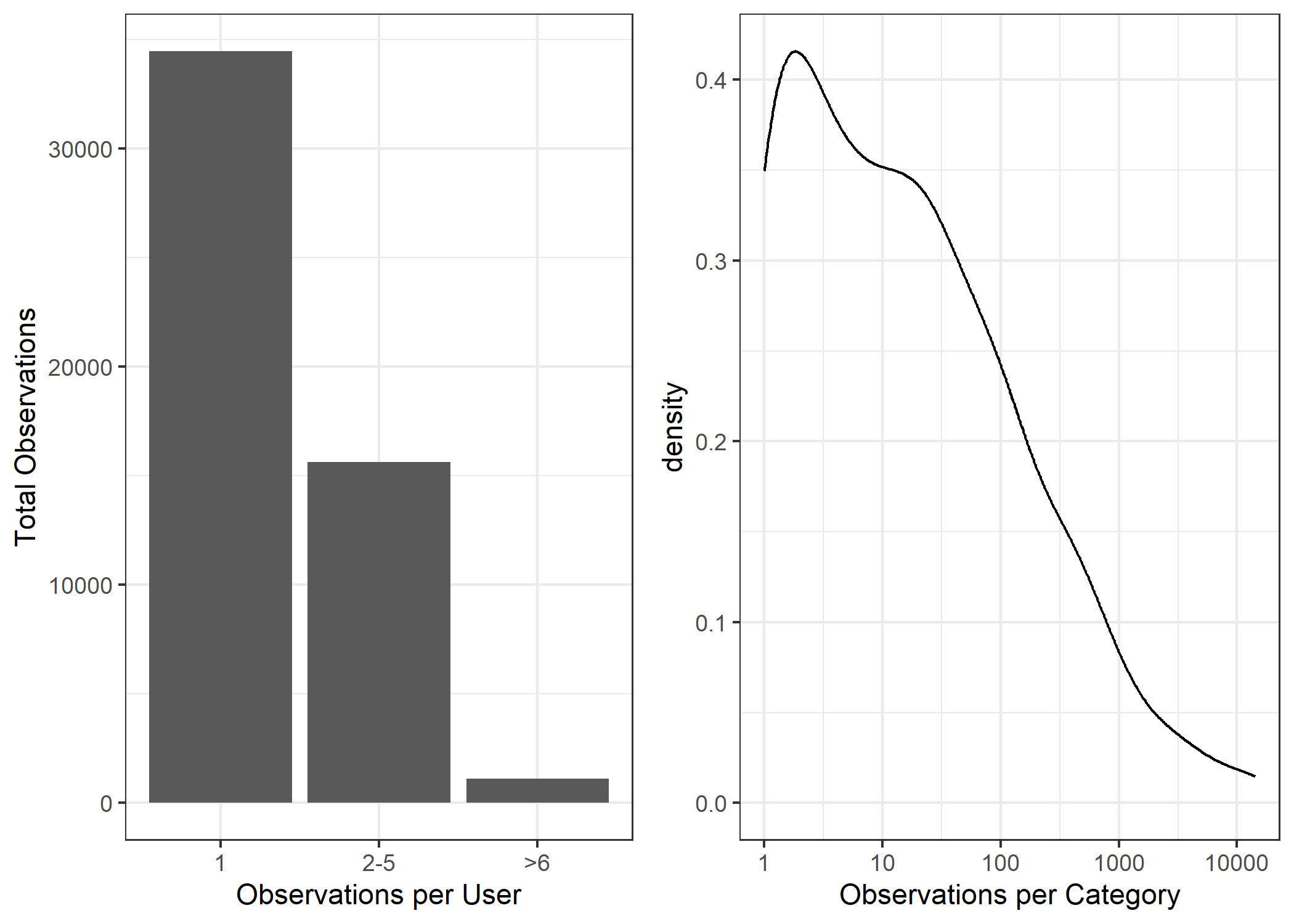}
    \caption{Observations per group for users (left) and tags (right). Most users have only a single observation. Tags, on the other hand, follow a right-tailed distribution.}
    \label{fig:obs_per_group}
\end{figure}

Part of this may be because of the relatively few observations per user (see fig \ref{fig:obs_per_group}). This highlights another issue with measuring Barrier-to-Exit across time: because it requires many observations over a long period of time to calculate sufficient precision of revealed preferences, it's hard to measure changes on the individual level.

While the effects of non-normality on the fixed effects should be minor \shortcite{schielzeth_robustness_2020}, we nevertheless conduct a robustness check. In the robustness check, we remove the problematic categories and refit the main model (eq. \ref{eq:model-equation}). The results of this can be seen in Appendix \ref{tricky_removed}

\section {Other Models}
\subsection {Model without activity-level} \label{no-activity}
The initial model, we had planned to test was a model without activity level ($\beta_2$), but otherwise following eq. \ref{eq:model-equation}. That is, all parameters have the exact same definition. For completeness, the old equation can be seen in eq. \ref{eq:old-model-equation}:

\begin{equation} \label{eq:old-model-equation}
    y_{ij} = \beta_{0} + \beta_{1} X_{1i} + e_{i} + e_{j} 
\end{equation}

However, when we ran the model (also using lmerTest \shortcite{kuznetsova_lmertest_2017}) and plotted the residuals, we got the following:

\begin{figure}[H]
    \centering
    \includegraphics[width=0.8\textwidth]{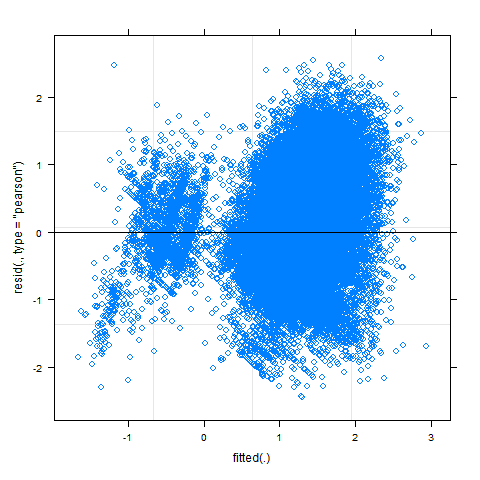}
    \caption{Residuals of the initial model. From a visual inspection, the residuals are \textit{not} randomly distributed}
    \label{fig:old_residuals}
\end{figure}

Just from a brief visual inspection, it is clear to see that the residuals are \emph{not} randomly distributed: There are two distinct "bands" that both seem to trend upward. This breaks the assumption that the residuals are randomly distributed \shortcite{poole_assumptions_1971}. While many assumption violations are reduced with enough data \shortcite{baayen_mixed-effects_2008,schielzeth_robustness_2020}, non-linearity is not one of them \shortcite{poole_assumptions_1971}. 

Fortunately, the non-random residuals were (partially) fixed by introducing activity-level for the reasons described in section \ref{section:model}.

\subsection {Problematic Categories Removed} \label{tricky_removed}
Here we fit the main model (eq. \ref{eq:model-equation}, with problematic categories removed. We define a problematic category as a category with a fitted random effect of less than -0.5. We obtain this threshold by visually inspecting Fig. \ref{fig:qq_main_random}. 

The results of this fit can be seen below in \ref{table:problematic-removed}:

\begin{table}[!htbp]   
  \centering
  \caption{Ablation with problematic categories removed}
  \label{table:problematic-removed}
  \begin{tabular}{lc}
    \toprule
    & \textit{Dependent variable:} \\
    \cmidrule(lr){2-2}
    & Barrier-to-Exit (Log) \\
    \midrule
    Time (years) & 0.018$^{***}$ \\
    & (0.001) \\
    \addlinespace
    Activity-level (Log) & 0.619$^{***}$ \\
    & (0.001) \\
    \addlinespace
    Intercept & 0.321$^{***}$ \\
    & (0.016) \\
    \midrule
    Observations & 77,058 \\
    Unique users & 46,668 \\
    Unique categories & 322 \\
    \midrule
    Marginal $R^2_{LMM}$ & 0.70 \\
    Conditional $R^2_{LMM}$ & 0.81 \\
    \bottomrule
    \multicolumn{2}{r}{\textit{Note:} $^{*}$p$<$0.1; $^{**}$p$<$0.05; $^{***}$p$<$0.01} \\
  \end{tabular}
\end{table}

The most important finding from this is that there is no difference in our variable of interest (Time in years). It is still 0.018 with a high degree of significance ($p \ll 0.0001$). Also worth noting is that the marginal $R^2$ has increased somewhat ($0.64 \rightarrow 0.70$) and the conditional $R^2$ is roughly the same (0.82 vs 0.81).

\subsection{Gamma mixed-effects model} \label{gamma model}
In the following, we refit the main model (eq. \ref{eq:model-equation}) using a Gamma regression. This is the most widely recommended solution in the literature on fitting right-tailed, heteroscedastic outcomes \shortcite{feng_log-transformation_2014,villadsen_statistical_2021}. However, since we discovered this after running our initial models, we could only justify doing this as a \textit{post-hoc} test.

I use the \texttt{lme4}-package to fit the model \shortcite{bates_fitting_2014}. To avoid convergence errors and adapt the model to the formula, we make the following alterations: 

\begin{enumerate}
    \item Add a gamma log-link function \shortcite{fox_applied_2015}
    \item change the year ($\beta_1$) estimate to decades. This has the effect of rescaling the effect size. 
    \item We still log-transform the activity-level to rescale it. As this is not part of the hypothesis, this does not affect our interpretation. 
\end{enumerate}

\begin{table}[!htbp] 
  \centering 
  \caption{Results of Gamma GLMM} 
  \label{table:gamma_results} 
  \begin{tabular}{lc} 
    \toprule
    & \textit{Dependent variable:} \\
    \cmidrule(lr){2-2}
    & Barrier-to-Exit (log) \\ 
    \midrule
    Time (decades) & 0.31$^{***}$ \\ 
    & (0.006) \\ 
    \addlinespace
    Ratings in period (Log) & 0.63$^{***}$ \\ 
    & (0.001) \\ 
    \addlinespace
    Intercept & 0.022$^{***}$ \\ 
    & (0.024) \\ 
    \midrule
    Unique Users (VPC) & 51,208 (0.24) \\
    Unique categories (VPC) & 330 (0.34) \\
    Observations & 84,806 \\
    \midrule
    Marginal $R^2$ & 0.64 \\
    Conditional $R^2$ & 0.91 \\
    \bottomrule
    \multicolumn{2}{r}{\textit{Note:} $^{*}$p$<$0.1; $^{**}$p$<$0.05; $^{***}$p$<$0.01} \\
  \end{tabular} 
\end{table}

\begin{figure}[H]
    \centering
    \includegraphics{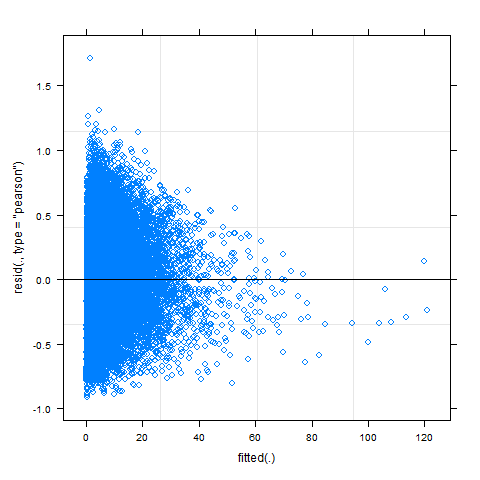}
    \caption{Residuals for the Gamma GLM. The residuals are heteroscedastic and have visible non-randomness.}
    \label{fig:glme_residuals}
\end{figure}

The results can be seen in table \ref{table:gamma_results}. Before interpreting the results there are a couple of things to note. For one, the marginal and conditional $R^2_{GLMM}$ using the tri-gamma method as recommended by \shortciteA{bates_fitting_2014}. Second, the residuals appear heavily non-random on a D\ref{fig:glme_residuals}. This might be because there are relatively few observations with high values of Barrier-to-Exit given its right-tailed nature or the unidentifiable fit. We therefore cannot make any strong interpretations from the data.

This leads us to the interpretation. The effect size per decade is 0.31, which is highly significant (SE=0.006, T=24, $p \ll 0.001$). This translates into 34\% increase per decade or 3\% increase per year. This is the same direction as our transformed linear model, with somewhat larger results. However, as these come from an almost unidentifiable fit with extremely non-random residuals, no inferences can be drawn from this.

Some of this indicates that the conditional distribution of Barrier-to-Exit is not a Gamma distribution. Pursuing the GLMM path would require further assessments of the best-fitting distribution. This could be by e.g. applying the Box-Cox method as described by \shortciteA{villadsen_statistical_2021}.

\section{Pre-processing steps} \label{preprocessing}
Dealing with a dataset with millions of rows and complex types like "categories" and "dates" requires special engineering considerations. This section outlines the pre-processing steps required to get the data from \shortciteA{ni_justifying_2019} in an analysis-ready shape.

All pre-processing of the data was done using python \shortcite{van_rossum_python_2007}. This is particularly because of the rich ecosystem of scientific packages. For this project we use \textit{numpy} \shortcite{harris_array_2020}, \textit{pandas} \shortcite{mckinney_pandas_2011}, and \textit{numba} \shortcite{lam_numba_2015} for efficient large-scale data processing. We also use \textit{scikit-learn} \shortcite{pedregosa_scikit-learn_2011} to efficiently parse categories (see \href{https://github.com/Rysias/amazon-book-coercion}{repository} for implementation). 

Most computations were performed on the Oxford Internet Institute's HPC cluster. This allowed us to benefit from multi-core processing \shortcite{gorelick_high_2020} and increased RAM.

The first step is creating a dataset of category relevance for the books (see \href{https://github.com/Rysias/amazon-book-coercion}{repository} for details). Here, we simply take the original gzipped file and extract a list of categories and item ID (\textit{asin}). This drastically reduces the file size, so we can do the computations in memory.

The next step is preparing the rating data. We start by filtering the dataset to only have users with more than 20 ratings. This reduces the dataset considerably as we saw in Fig. \ref{fig:people}. We then left-join the data with the category similarity data described above. Each row now consists of a $user\_id$, $category\_id$, $timestamp$, and $preference\_score$ (i.e. the rating multiplied by category relevance; see eq. \ref{eq:preference}) for each rating that the user has made for any given category. Note, that each individual rating can be represented in multiple rows if a book has multiple categories (which most have).

Finally, we summarise the data to get the sum of preference scores and amount of ratings per user, category, and quarter. This gives us a further reduced dataset that is more manageable to work with. Revealed preferences are defined as the weighted sum of ratings and category relevance (eq. \ref{eq:preference}), so this decision is mainly one of granularity. 

\end{document}